\documentstyle[12pt]{article}

\title{\begin{flushright}
{\normalsize
CERN-TH/99-228\\
NUC-MINN-99/12-T\\
July 1999 \\}
\end{flushright}
\vspace*{0.3in}
{\bf NEUTRAL KAON SYSTEM IN DENSE MATTER AND HEAVY-ION COLLISIONS}}
\author{{\bf Giovanni AMELINO-CAMELIA}$^{\dag}$ and {\bf Joseph
KAPUSTA}$^{\dag,\ddag}$
 \vspace*{0.1in}\\
 {\it $^{\dag}$Theory Division, CERN}\\ \vspace*{0.2in}
 {\it CH-1211 Geneva 23, Switzerland}\\ 
  {\it $^{\ddag}$School of Physics and Astronomy, University of Minnesota}\\
  {\it Minneapolis, MN 55455, USA}}

\date{}

\begin{document}

\maketitle

\begin{center}
Abstract
\end{center}

Above a critical matter density the propagating modes of the neutral
kaon system
are essentially eigenstates of strangeness, but below it they are almost
complete eigenstates of CP.  We estimate the real and imaginary parts of the
energies of these modes and their mixing at all densities up to
nuclear matter density $2\times 10^{14}$ g/cm$^3$.  In a heavy ion
collision the
strong interactions create eigenstates of strangeness, and these propagate
adiabatically until the density has fallen to the critical value,
whereupon the
system undergoes a sudden transition to (near) eigenstates of CP.
We estimate the critical density to be 20 g/cm$^3$, and that this
density will be reached
about $2\times 10^5$ fm/c after the end of the collision.

\newpage

Neutral kaon systems are extremely interesting for a variety of reasons
\cite{texts1}.  When they are created in a collision among hadrons or leptons
they are essentially in eigenstates of strangeness because the strong and
electromagnetic interactions conserve flavor.  When they propagate freely in
vacuum the weak interactions are operative and both C and P are violated.  The
eigenstates of the full Hamiltonian are then almost completely flavor mixed as
short- and long-lived kaons:
\begin{eqnarray}
|K_S\rangle &=& \left[ (1+\epsilon)|K^0\rangle - (1-
\epsilon)|\bar{K}^0\rangle\right]
/\sqrt{2(1+|\epsilon|^2} \nonumber \\
|K_L\rangle &=& \left[ (1+\epsilon)|K^0\rangle + (1-
\epsilon)|\bar{K}^0\rangle\right]
/\sqrt{2(1+|\epsilon|^2} \, .
\end{eqnarray}
Here $|\epsilon| \approx 2\times 10^{-3}$ is the measure of CP violation.  When
a beam of long-lived kaons is sent through ordinary matter, short-lived kaons
are generated due to the different interactions between the components of the
former, namely $K^0$ and $\bar{K}^0$, and atomic nuclei.  This is called kaon
regeneration.  The goal of this paper is to study the collective modes of
propagation of the neutral kaons in {\it dense} matter, and to determine the
fate of these modes after they are created in a collision between large nuclei
at high energy.

We will use a strong interaction basis with
\begin{eqnarray}
|K^0\rangle &=& | d\bar{s}\rangle = |1\rangle = \left( \begin{array}{c}
1 \\ 0 \end{array} \right) \nonumber \\
|\bar{K}^0\rangle &=& |\bar{d}s \rangle = |2\rangle = \left( \begin{array}{c}
0 \\ 1 \end{array} \right) \, .
\end{eqnarray}
Because it is so small and plays no special role in our analysis we shall set
$\epsilon = 0$.  Then $|K_S\rangle$ and $|K_L\rangle$ are eigenstates of
CP with eigenvalues $+1$ and $-1$, respectively.

First we do a relativistic analysis.  Poles of the propagator determine the
time
evolution of small amplitude excitations.  These are obtained from solutions
to the equation
\begin{equation}
\omega^2 -k^2 -\Pi_{vac}(\omega^2-k^2) - \Pi_{mat}(\omega,k) = 0 \, ,
\end{equation}
where the vacuum and matter contributions to the self-energy are indicated.
Generally it is an excellent approximation to evaluate these $2\times2$
matrices
on the mass shell.  This means that $\Pi_{vac}$ is a constant and $\Pi_{mat}$
depends on the momentum $k$ only.  The usual analysis gives
\begin{equation}
\Pi_{vac} = \left( \begin{array}{cc} A & B \\ B & A \end{array} \right)^2
\end{equation}
where $A$ and $B$ are complex numbers.  In terms of measurables they are
\cite{PDG}:
\begin{eqnarray}
A &=& m_K -i(\Gamma_S+\Gamma_L)/4 \nonumber \\
B &=& \Delta m/2 +i\Delta \Gamma \nonumber \\
m_K &=& (m_S+m_L)/2 = 497.67 \; {\rm MeV} \nonumber \\
\Delta m &=& 3.52\times10^{-12} \; {\rm MeV} \nonumber \\
\Delta \Gamma &=& \Gamma_S-\Gamma_L \nonumber \\
\Gamma_S &=& 7.38\times10^{-12} \; {\rm MeV} \nonumber \\
\Gamma_L &=& 1.27\times10^{-14} \; {\rm MeV} \nonumber \\
\Delta m &=& (0.478\pm0.003) \Delta \Gamma \, .
\end{eqnarray}
The matter contribution is diagonal in flavor and is expressed as
\begin{equation}
\Pi_{mat} = \left( \begin{array}{cc} F & 0 \\ 0 & \bar{F} \end{array}
\right) \, .
\end{equation}
In matter with an excess of baryons over antibaryons the strange and
antistrange
components, $F$ and $\bar{F}$, are different.  Of course, this is the origin
of kaon regeneration in matter.  One reason they are different is because the
valence antiquark $\bar{d}$ in the $\bar{K}^0$ can annihilate on a valence
quark $d$ in the proton or neutron, producing a hyperon.  This is not
possible with a $K^0$.

Diagonalization of the equation
\begin{equation}
(k^2 + \Pi) \left( \begin{array}{cc} \alpha \\ \beta \end{array} \right) =
\omega^2 \left( \begin{array}{cc} \alpha \\ \beta \end{array} \right)
\end{equation}
yields the energy eigenvalues
\begin{equation}
\omega^2_{\pm} = k^2 + A^2 + B^2 + \frac{1}{2} (F+\bar{F})
\pm \sqrt{4A^2B^2 + \frac{1}{4} (F-\bar{F})^2}
\end{equation}
and eigenstates
\begin{equation}
\beta_{\pm} = \left( -\chi \pm \sqrt{1+\chi^2} \right) \alpha_{\pm}
\end{equation}
where $\chi = (F-\bar{F})/4AB$. The states evolve in time as $\exp(-
i\omega_{\pm}t)$.  In baryon-free matter, even at finite temperature, there is
no difference between $F$ and $\bar{F}$ (from the point of view of the strong
and electromagnetic interactions).  Then, with the above conventions,
$\beta_{\pm} = \pm \alpha_{\pm}$, and the upper sign corresponds to $K_L$ and
the lower sign
to $K_S$.  These are the familiar modes in vacuum except that the square of the
mass is shifted by $F$, which generally has both real and imaginary
parts.

At nuclear matter density, 0.155 nucleons/fm$^3$, the situation is very
different.  One should expect that $|F-\bar{F}|$ is only somewhat smaller than
typical nuclear energies, implying something on the order of (100 MeV)$^2$.
Since
\begin{equation}
2AB = m_K(\Delta m + i\Delta \Gamma/2) = (4.19\times10^{-5} \; {\rm MeV}
)^2 + i(4.28\times10^{-5} \; {\rm MeV})^2
\end{equation}
this means that $|\chi| >> 1$.  In this case, the off-diagonal matrix elements
in the self-energy,
due to the weak interactions, are totally ignorable, and the
energy eigenstates are also eigenstates of strangeness.

Let us now estimate $F-\bar{F}$ as a function of density.  At low to moderate
densities the self-energy may be expressed in terms of the scattering amplitude
for kaons scattering on the constituents of the medium, here taken to be an
equal mixture of protons
and neutrons \cite{selfe}.
\begin{eqnarray}
F(\omega,k) &=& - 4\pi \int \frac{d^3q}{(2\pi)^3} \,
4 n_{FD}(q) \, \sqrt{\frac{s}{m_N^2+q^2}}
 \, f(s) \nonumber \\
&=& -4\pi \rho_N \left(1+\frac{m_K}{m_N} \right) \langle f \rangle \, .
\end{eqnarray}
In the first line, $n_{FD}(q)$ is the Fermi-Dirac distribution for a nucleon
with momentum $q$, the extra factor of 4 appears because of summation over spin
and isospin of the nucleons, $f$ is the forward scattering amplitude in the
center-of-momentum frame, and $s$ is the usual Mandelstam variable.  The angular
brackets in the second line denote the particular momentum averaging, and
$\rho_N$ is the spatial density of nucleons.  If the nucleons are not
distributed according to a Fermi-Dirac distribution then one ought to use the
actual momentum distribution; this will affect the numerical value of $\langle f
\rangle$ to some extent.  Formula (11) is simply a version of the low density
virial expansion familiar in statistical physics.  Its applicability has been
discussed in \cite{selfe}.  Essentially it requires that the average separation
of nucleons be larger than both the real part of the forward scattering
amplitude and the inverse of the average relative momentum between a nucleon and
a kaon.  One may think of this formula quantum mechanically as representing
indices of refraction (real part) and attentuation (imaginary part).  Indeed, it
is only a variation of the usual formulae used in kaon regeneration sudies.

There is an identical expression
for antikaons with $\bar{F}$ replacing $F$ and $\bar{f}$ replacing $f$.  The
difference $\bar{f}-f$ has been estimated, or may be inferred, from several
sources.  For example, Eberhard and Uchiyama \cite{EU} calculated the real and
imaginary parts of the forward scattering amplitudes for neutral kaons incident
on proton and nuclear targets.  These calculations were based on measured total
cross sections for $K^+$ and $K^-$ on protons and neutrons, on published values
of $K$-nucleon elastic scattering, and charge symmetry.  From their figures 2
and 3 we estimate that
\begin{displaymath}
\langle \bar{f}-f \rangle \approx 0.3(1+i) \; {\rm fm} \, .
\end{displaymath}
Here the averaging is for kaon momenta ranging from 0.3 to 1 GeV, relevant for
the conditions in hot and dense nuclear matter created in a heavy
ion collision.
Shuryak and Thorsson \cite{ST} have calculated the scattering amplitudes for
charged kaons incident on nucleons based on partial
wave analyses of data. From their figures 1 and 2 we estimate that
\begin{displaymath}
{\rm Re}\, \langle \bar{f}-f \rangle \approx 1.5 \; {\rm fm} \, .
\end{displaymath}
Here the averaging is for $\sqrt{s}$ ranging between 1.5 to 1.8 GeV
corresponding to the range of kaon momenta quoted above.  (Unfortunately they
have only displayed the real parts explicitly.)  We do not know the reason for
this discrepancy.  Fortunately the precise numbers are totally irrelevant for
the mixing of the neutral kaons in dense matter, since even the smallest
estimate of the strong interaction induced self-energy overwhelms that due to
the weak interactions.  For example, using the estimate from Eberhard and
Uchiyama, we obtain
\begin{equation}
F-\bar{F} \approx  (150 \; {\rm MeV})^2 (1+i) \left( \frac{\rho_N}{0.1 / {\rm
fm}^3}\right) \, ,
\end{equation}
which ought to be compared to eq. (10).  The fact that the real part of $F-
\bar{F}$ is positive implies that the nonrelativistic one-body potential, $U =
F/2m_K$, is relatively more repulsive for kaons than for antikaons.  This is
quite natural for the reasons stated earlier.  The fact that the imaginary part
is positive is also to be expected because the imaginary part of the scattering
amplitude and cross section are related by the optical theorem, $\sigma =
(4\pi/k_{cm}){\rm Im}\, f$, and antikaons have a much bigger inelastic cross
section with nucleons than kaons.

An alternative approach to the kaon self-energy at moderate to high density is
to use an effective Lagrangian incorporating the relevant degrees of freedom and
symmetries \cite{KN}.  One finds that \cite{pie}
\begin{eqnarray}
F(\omega,k) &=& -\frac{\Sigma_{KN}}{f^2_{\pi}}\rho_S +
\frac{3}{4}\frac{\rho_N}{f_{\pi}^2} \omega \nonumber \\
\bar{F}(\omega,k) &=& -\frac{\Sigma_{KN}}{f^2_{\pi}}\rho_S -
\frac{3}{4}\frac{\rho_N}{f_{\pi}^2} \omega \, .
\end{eqnarray}
Here $\Sigma_{KN}$ is the kaon-nucleon sigma term, estimated to be of order 350
MeV.  One distinquishes the scalar nucleon density, $\rho_S
= \langle \bar{N}N \rangle$, from the conserved vector
density, $\rho_N = \langle \bar{N} \gamma^0 N \rangle$,
although they only begin to differ significantly above twice nuclear density.
The difference in sign in the above equations results from the fact that the
scalar density treats particles and antiparticles the same whereas the vector
density distinguishes particles from antiparticles.  Thus
\begin{equation}
{\rm Re}\, (F-\bar{F}) = \frac{3}{2}\frac{\rho_N}{f_{\pi}^2} \omega =  (260 \,
{\rm MeV})^2 \left(\frac{\omega}{m_K}\right) \left(\frac{\rho_N}{0.1 / {\rm
fm}^3}\right)
\end{equation}
which lies between the estimates from Eberhard-Uchiyama and Shuryak-Thorsson.
So far, to our knowledge, no one has calculated the imaginary part in this
approach.

What happens to the neutral kaons after production in a heavy ion collision?
Based on the above analysis of $F-\bar{F}$ we see that the dimensionless ratio
$\chi$ is proportional to the nucleon density and has a very small imaginary
part relative to the positive real part.
Using the results of Eberhard-Uchiyama we get
\begin{equation}
\chi = 6.3\times10^{12} \, \left(\frac{\rho_N}{0.1 / {\rm fm}^3}\right) \, .
\end{equation}
From eq. (9) one finds that $|\alpha_+| \gg |\beta_+|$ and $|\beta_-| \gg
|\alpha_-|$.  Hence the eigenstates
are $|+\rangle = |1\rangle$ and $|-\rangle =
|2\rangle$ with corrections of order $1/\chi$.  As the matter expands the
density decreases until $\chi$ becomes very small.
Then the eigenstates are the
vacuum ones, $|K_L\rangle$ and $|K_S\rangle$.  {\em If} this evolution is
adiabatic it would have extremely interesting consequences.
A $K^0$ produced in
the collision would evolve into a $K_L$ (rather than into an equal mixture of
$K_L$ and $K_S$), and a $\bar{K}^0$ would evolve into a $K_S$.
Based on valence quark counting \cite{chem} one expects that
\begin{equation}
\left( \frac{K_L}{K_S} \right)_{\rm observed} =
\left( \frac{K^0}{\bar{K}^0} \right)_{\rm in \; matter} =
\left( \frac{K^+}{K^-} \right) \left( \frac{\pi^-}{\pi^+} \right) \, .
\end{equation}
That is, the observed ratio of $K_L$ to $K_S$ should be equal to the ratio of
$K^0$ to $\bar{K}^0$ produced in dense matter, which then is equal to the ratio
of charged kaons times the ratio of charged pions.
The latter product of ratios
should not change during the very late dilute expansion phase of a heavy ion
collision.  It has been measured \cite{AGS} in central Au-Au collisions at
the AGS at Brookhaven National Laboratory (E$_{\rm beam}$ = 11 GeV/nucleon)
with the value 6.  It has also been measured \cite{SPS} in central Pb-Pb
collisions at
the SPS at CERN (E$_{\rm beam}$ = 160 GeV/nucleon) with the value 2.  Hence,
contrary to all other high energy accelerator experiments, the ratio of
long-lived to short-lived kaons would be far from one!  But does the neutral
kaon system evolve adiabatically?

To answer this question, consider what happens in the kaon's own frame of
reference.  As time goes on, the surrounding density of matter decreases as
$1/t^3$.  This naturally follows from dimensional analysis, and it also emerges
from a calculation of the free expansion of an ideal relativistic gas
\cite{Dirk}.  A good estimate of the local nucleon density is $\rho_N(t) =
\rho_0 (t_0/t)^3$.  For numerical estimates we shall use $\rho_0 = 0.1$
nucleons/fm$^3$ and $t_0 = 10$ fm/c, which are characteristic of central
collisions between nuclei of atomic number near 200 at the AGS and the SPS
\cite{qm}.  Furthermore, the average relative speed in
an encounter between a kaon and a nucleon will decrease with time.  The reason
is that the fireball has an initial radius of $R \approx 10$ fm or so.   After
an elapsed time $\Delta t = t-t_0$ nucleons with a relative speed greater than
$R/\Delta t$ are unlikely to ever encounter the kaon.  If we are interested in
what happens at late times in the expansion, when the strong and weak
interactions affecting the neutral kaon system become comparable, the relevant
averaged scattering amplitudes are not those discussed above.  Rather, one can
use only the s-wave scattering lengths.  A compilation of data yields \cite{a0}
\begin{equation}
\bar{f}_0 - f_0 = 0.1 + 0.6 i \; {\rm fm} \, ,
\end{equation}
with an uncertainty of about 0.1 fm in both the real and imaginary parts.
This all leads to the difference of one-body potentials being
\begin{equation}
U(t) - \bar{U}(t)
= (7.5+45i)\left( \frac{10 \; {\rm fm/c}}{t} \right)^3 \; {\rm MeV} \, .
\end{equation}

The problem can be reduced to a two-level Schr\"odinger equation with a
time-dependent complex potential.  Taking out the kaon mass leads to
\begin{equation}
i\frac{\partial}{\partial t} \psi(t) = H(t) \psi(t) = \left( \begin{array}{cc}
-\frac{i}{4}(\Gamma_S + \Gamma_L) + U(t) &
\frac{1}{2}\Delta m + \frac{i}{4}\Delta \Gamma \\
\frac{1}{2}\Delta m + \frac{i}{4}\Delta \Gamma &
-\frac{i}{4}(\Gamma_S + \Gamma_L) + \bar{U}(t) \end{array}
\right) \psi(t) \, .
\end{equation}
The instantaneous energy eigenvalues are
\begin{equation}
E_{\pm}=\frac{1}{2}(U+\bar{U}) -\frac{i}{4}(\Gamma_S + \Gamma_L)
\pm \sqrt{\left(\frac{1}{2}\Delta m + \frac{i}{4}\Delta \Gamma \right)^2
+ \frac{1}{4} (U-\bar{U})^2} \, .
\end{equation}
The instantaneous eigenstates are given by eq. (9).  The exact solutions can be
expanded in terms of these, with time-dependent coefficients $a_{\pm}(t)$, as
\begin{equation}
\psi(t) = a_+(t) |+,t\rangle + a_-(t) |-,t\rangle
\end{equation}
where
\begin{equation}
|\pm,t \rangle = \left( \begin{array}{cc} \alpha_{\pm}(t) \\ \beta_{\pm}(t)
\end{array} \right) \exp\left\{ -i\int_0^t E_{\pm}(t') dt' \right\} \, .
\end{equation}
There exists a critical density $\rho_c$ defined by the condition
that $|\chi| = 1$.  With the above input its numerical
value is $\rho_c = 1.1 \times 10^{-14}$
nucleons/fm$^3$ or about 19 g/cm$^3$.  For densities much greater
than $\rho_c$
the strong interactions dominate and the eigenstates of the system are
eigenstates of strangeness.  For densities much less than $\rho_c$ the weak
interactions dominate and the eigenstates of the system are eigenstates of CP.
The critical region may be conservatively defined as $8 > |\chi| > 1/8$
or $8 > \rho_N/\rho_c > 1/8$.

The picture that emerges is as follows.  For $t < 10^5$ fm/c the matter
expands freely, with frequent
(on the time scale of $1/\Delta m$ and $1/\Delta \Gamma$)
interactions of the kaons with the nucleons maintaining
the kaons in eigenstates
of strangeness, namely, $K^0$ and $\bar{K}^0$.  For $10^5 < t < 4\times 10^5$
fm/c the system is in a transition region where $|\chi| \approx 1$,
meaning that interactions with the nucleons
are comparable in strength with the internal weak
interactions of the kaons.  In this region there is nothing to prevent
transitions between the states.  For $t > 4\times 10^5$ fm/c the matter is so
dilute that the internal weak interactions of the kaons dominate and they
propagate essentially as in vacuum, namely, as $K_L$ and $K_S$.
Thus the states
of the neutral kaon system evolve adiabatically at both high and low density.
The time to pass through the transition region, about $3\times 10^5$ fm/c,
is so short compared to
the natural oscillation time of the neutral kaons, $1/\Delta m
= 5.6\times 10^{13}$ fm/c, that this transition may be treated with the sudden
approximation.  It is at the time $t \approx 2\times 10^5$ fm/c and the density
$\rho_c \approx 20$ g/cm$^3$ that the strangeness eigenstates $K^0$ and
$\bar{K}^0$ decompose into the eigenstates $K_L$ and $K_S$ of CP.  This may be
shown mathematically from the equations of motion of
$a_{\pm}$.\footnote{However, the analysis is more
complicated than that given in
textbook discussions of time-dependent
perturbation theory and the adiabatic and
sudden approximations because the Hamiltonian is not Hermitian, and the
instantaneous eigenstates are not orthogonal in the transition region.}
As a consequence of this sudden transition the observed ratio of
long-lived to short-lived kaons should be 1.
Finally we should remark that the
probability for a kaon to decay before the transition
region is reached is negligible because the lifetime of even $K_S$ is much
greater than the time to reach the critical density.

The phenomenon described here does not happen in elementary particle collisions
such as $e^+e^-$, $p\bar{p}$, and $pp$ because the net baryon number is either
zero or negligibly small.  It should be noted that the critical density of 20
g/cm$^3$ is characteristic of heavy metals\footnote{In this case $f-\bar{f}$ is
the difference of scattering amplitudes on a nucleus and $\rho_N$ is replaced by
the density of these nuclei}, perhaps opening the window on new
types of experiments with neutral kaons.

\section*{Acknowledgements}

We thank U. Heinz and P. J. Ellis for useful conversations.
J.K. thanks the Institute of Technology at the University of Minnesota for
granting a single quarter leave in the spring of 1999 and
the Theory Division at
CERN for hospitality and support during that time.  This work was supported by
the US Department of Energy under grant DE-FG02-87ER40328 and by
the European Union under a TMR grant.

\end{document}